\documentclass[runningheads]{llncs}
\usepackage{makeidx}  % allows for indexgeneration
\usepackage{graphicx}
\usepackage{subfigure}
\usepackage{amssymb,amsmath,bm}
\usepackage{mathrsfs}
\usepackage{booktabs}
\usepackage{amsmath}
\usepackage{mathtools}
\usepackage{subfig}
\usepackage[misc]{ifsym}
\usepackage{multirow}
\usepackage{bbding}
\usepackage{array}
\usepackage{float}
\usepackage{algorithm}
\usepackage{algorithmic}
\usepackage{lettrine}
\usepackage{cite}

\newcolumntype{C}[1]{>{\centering\let\newline\\\arraybackslash}m{#1}}
\renewcommand\arraystretch{1.5}
\usepackage{pbox}
\usepackage{gensymb}

\begin{document}
	\frontmatter          % for the preliminaries
	\title{Ensembled ResUnet for Anatomical Brain Barriers Segmentation}
	
%	\author{*******}
%	
%	\institute{***}
	
	\author{Munan Ning$^{1}$ \and Cheng Bian$^{1}$ \and Chenglang Yuan$^{1}$ \and Kai Ma$^{1}$ \and Yefeng Zheng$^{1}$}
	\institute{Tencent Jarvis Lab \\
		\email{\{masonning, tronbian, nikoyuan, kylekma, yefengzheng\}@tencent.com}}
	
	\maketitle
	
	\begin{abstract}
    Accuracy segmentation of brain structures could be helpful for glioma and radiotherapy planning. However, due to the visual and anatomical differences between different modalities, the accurate segmentation of brain structures becomes challenging.
    To address this problem, we first construct a residual block based U-shape network with a deep encoder and shallow decoder, which can trade off the framework performance and efficiency. Then, we introduce the Tversky loss to address the issue of the class imbalance between different foreground and the background classes. Finally, a model ensemble strategy is utilized to remove outliers and further boost performance.
	\end{abstract}
	
	\section{Introduction}
	Gliomas are the most common type of brain tumors, originating from glial cells in the human brain. In clinical radiation therapy, the precise identification of the clinical target volume (CTV) boundary can ensure the operation effect, whereas the CTV boundary is determined by anatomical structures. Therefore, the accurate and automatic segmentation of brain anatomical structures could improve both efficiency and effectiveness for gliomas treatment.
	
    Recently, deep learning approaches have been proven its the superiority on both 2D natural images and 3D medical modalities segmentation task compared to the traditional methods. Especially for the brain tumor segmentation, ~\cite{myronenko20183d} proposed a U-Net with an additional VAE branch to reconstruct input images and regularize the shared encoder. ~\cite{jiang2019two} established a two-stage cascaded U-Net to refine the coarse prediction from the first stage and capture context information in the second stage. As to brain CTV segmentation, ~\cite{ermics2020fully} introduced DenseNet to predict resection cavity precise contours.
    
    To raise the researcher interest in the study of brain CTV segmentation, the Anatomical Brain Barriers to Cancer Spread challenge (ABCs) aims to encourage challengers to construct an automatic brain structures segmentation methods, where some critical structures (e.g., structures that are served as barriers to the spread of brain cancers or spared from irradiation) are included in two tasks.
    The dataset provides 45 multi-modal images with ground truth annotations for training and validation, 15 images for the online test, and 15 images for the final test. Each case is given with a CT scan and two diagnostic MRI scan which includes contrast-enhanced T1-weighted and T2-weighted FLAIR of the post-operative brain. Besides, all images are co-registered and re-sampled to the size of 164x194x142 pixels with the isotropic resolution of 1.2x1.2x1.2 mm. For the task 1, challengers are asked to segment brain structures for the automatic identification of the CTV. As for the task 2, participants are required to segment structures that used in radiotherapy treatment plan optimization. Evaluation metrics are specified to the Dice score and surface Dice score to evaluate the performance of the proposed algorithms.
    
    In this report, we propose a residual block~\cite{he2016deep} based U-Net which is composed of a deep encoder and a shallow decoder and implemented with nn-UNet framework. To provide multi-scale guidance for boosting framework performance and accelerate training convergence, we employ a deep supervision strategy in our framework. To suppress the false-negative results and retrieve the missing small targets, we utilize the Tversky loss together with the cross-entropy loss as our criterion. Experiments show that the proposed framework outperforms the compared state-of-the-art methods.

% 	Clinically, there are multiple ways of glaucoma measurement, which consist of measuring the Intraocular Pressure (IOP), performing visual field test performing ONH examination and calculating Cup-to-Disc Ratio (CDR) in fundus image.

    \section{Method} 
    Based on the U-Net ~\cite{ronneberger2015u} architecture and nn-UNet ~\cite{isensee2018nnu} framework, we propose a variant of this approach based on residual blocks. The details are illustrated as follows.
    \begin{figure}[ht]
	\centering
	\includegraphics[width=1.0\columnwidth]{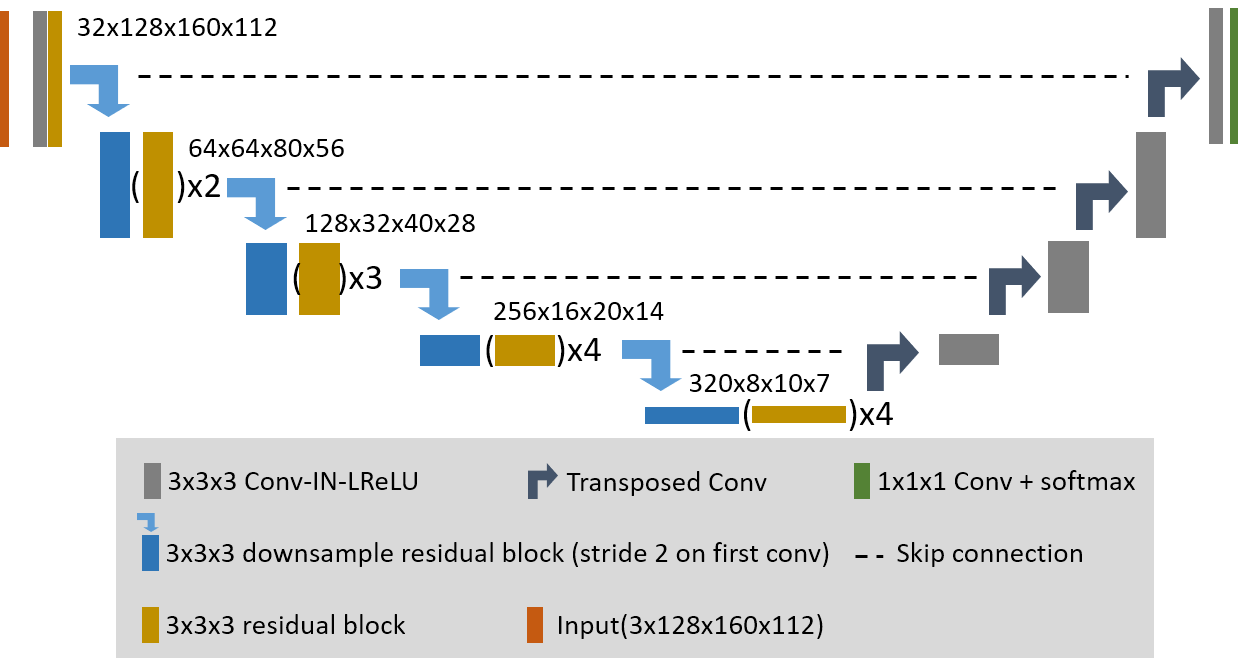}
	\caption{An overview of the proposed framework.}
	\label{fig:Framework}
    \end{figure}

	\subsection{Encoder design}
	The proposed encoder is composed of residual blocks as shown in Fig.1, which consist of two 3x3x3 3D convolutions layers following with a normalization layer and a activation layer. Then, an identity skip connection operation is employed to link the shallow and the corresponding deep level features. Specifically, we use Instance Normalization (IN) as the normalization layer, since it performs better than BatchNorm when batch size is small (ours is 2), and requires less computation cost than Group Normalization (GN). In addition, We utilize LeakyReLU as the activation layer to retain more information than ReLU. Totally, we have 5 spatial levels with the number of residual blocks in each level of 1, 2, 3, 4, 4. The number of filters, which is implemented by a 3D stridden convolution layer with 3x3x3 filter and a stride of 2, is initiated with 32. Then, we doubles this number after each downsample operation. Notably, the number of the filter in the last level is set to 320 for computational efficiency. The last downsample convolution avoid the axis z for retaining the slice-wise resolution. 
    We randomly crop the patch with the size of 3x128x160x112 from 3-modality images as the framework input. After the proposed framework processing, the final output is obtained with a size of 320x8x10x7.
	
	\subsection{Decoder design}
	The decoder structure is the opposite operation of the encoder, composed of an upsample block and a standard convolution block, where the upsample block is implemented with a 1x1x1 3D convolution layer and a 3D transpose convolution layer. The 3D convolution layer aims to reduce the number of features by a factor of 2, and the 3D transpose convolution layer is used for double the spatial dimension. The 
    standard convolution block consists of a simple 3x3x3 3D convolution layer followed with a Instance Normalization and a Leaky ReLU layers. At the end of the decoder, a 1x1x1 convolution layer with the output channels of target classes conjoins a softmax activation layer to get the final prediction. The shallow decoder could not only boost the speed of training and inference stage but also avoid overfitting. 

    To accelerate model convergence and provide multi-scale guidance, we also introduce a deepvision strategy, where other three levels of the features will be processed with an extra output block so as to obtain the additional predictions and supervised by the same loss function. The final loss will be weighted differently with respect to the importance of each level output. Here, we set the weights of 8/15, 4/15, 2/15, and 1/15 for the outputs from level 1, 2, 3, and 4 respectively.
    
    \subsection{Loss function}
    We utilize the DC-CE loss as our criterion, which is composed of origin linear Dice loss and cross-entropy loss:
    \begin{equation}
        L_{DCCE} = -\frac{\sum_{i=1}^{N}\sum_{c=1}^{C} p_{i, c} y_{i, c}+\epsilon}{\sum_{i=1}^{N}\sum_{c=1}^{C} p_{i, c}+y_{i, c}+\epsilon}
                   -\frac{1}{N}\frac{1}{C}\sum_{i=1}^{N}\sum_{c=1}^{C} y_{i, c} \log \left(p_{i, c}\right)    
    \end{equation}
    where $N$ is the total number of pixels, and $C$ denotes all classes. $p_{i, c}$ and $y_{i, c}$ represent the prediction and annotation of pixel $i$ and class $c$, respectively.
    The Tversky loss is also introduced to train our proposed framework, which can be formulated as:
    \begin{equation}
        L_{Tversky}=\frac{\sum_{i=1}^{N}\sum_{c=1}^{C} p_{i, c} y_{i, c}+\epsilon}{\sum_{i=1}^{N}\sum_{c=1}^{C} p_{i, c} y_{i, c}+\alpha \sum_{i=1}^{N}\sum_{c=1}^{C} p_{i, c} \bar{y}_{i, c}+\beta \sum_{i=1}^{N}\sum_{c=1}^{C} \bar{p}_{i, c} y_{i, c}+\epsilon}
    \end{equation}
    where $p_{i, c} \bar{y}_{i, c}$ denotes the false positive (FP) pixels; $bar{p}_{i, c} y_{i, c}$ denotes the false negative (FN) pixels. The importance of FP and FN is weighted by $\alpha$ and $\beta$, which are set to 0.3 and 0.7, respectively.
    
    \subsection{Pseudo training with model ensemble}
    For a further improvement of segmentation accuracy, we gradually employ the model ensemble strategy and pseudo training strategy. At first, we calculate the average of all selected models to get the prediction of the test volumes:
    \begin{equation}
        P^{*}=\frac{1}{N} \sum_{i=1}^{N} P_{i}
    \end{equation}
    The $P_{i}$ denotes the specific prediction of different models, while the $P^{*}$ denotes the final ensembled prediction. We find it more stable than any single prediction since the error-prone outliers are removed in the ensemble process.
    
    Based on the reliable predictions, we treat them as the pseudo labels for the test volumes. The final loss function for the model training can be described as follow:
    \begin{equation}
        L_{hybrid}(V, L)=L_{DCCE}(V, L)+L_{Tversky}(V, L)
    \end{equation}
    \begin{equation}
        L_{final}=L_{hybrid}\left(V_{s}, L_{s}\right)+L_{hybrid}\left(V_{t}, P_{t}^{*}\right)
    \end{equation}
    We define the sum of $L_{DCCE}$ and $L_{Tversky}$ as $L_{hybrid}$, and utilize the ensembled prediction $P^{*}$ as the pseudo label for test volumes. 
    The pseudo training strategy can bring the model more training pairs, and the model can learn reliable information from pseudo labels. The result in Table ~\ref{table:t1} proved the effectiveness of the methods in the section.
    
    \subsection{Optimization}
    We utilize the SGD with initial learning rate of $\alpha_{0}=1e-4$ and momentum of 0.99 to optimize our network. Poly strategy is employed to progressively decrease learning rate according to:
    \begin{equation}
        \alpha=\alpha_{0} *\left(1-\frac{e}{N_{e}}\right)^{0.9}
    \end{equation}
    where $e$ an iterator of the current epoch; $N_e$ is the total number of the training epochs. We set $N_e=1000$ with batch size of 2 in the training stage. L2 norm regularization with a weight of $1e-5$ is used to prevent overfitting. 
    
	\section{Experiments} 
	\subsection{Implementation details and data processing} 
	In this section, we will demonstrate the experimental details of our method. Firstly, we adopt the normalization for three modalities images by subtracting the mean values and dividing the variances.
    Specifically, to reduce the negative effect from extremum in CT images, the gray intensity in CT modality is clipped into [0.5, 0.995].
    
    Afterwards, we concatenate 3 modalities to obtain a 3-channel volume as the framework input. A series of data augmentation strategies are employed, including rotation, scale, mirror, gamma correction, and brightness additive, to improve the robustness of the framework. As to the inference phase, the test time augmentation strategy(e.g., sliding window and flipping across three axes) is introduced to improve the performance. It is worth noting that adopting horizontal flip along with the x-axis could greatly reduce the accuracy in our experiment. The reason might be that such operation will misleads the framework when learning the paired tissues such as eyes and cochlea. Therefore, we decide to avoid the horizontal flip along with x-axis in Task 2. To evaluate the effectiveness of the proposed framework, we employ the 5-fold cross validations, and choose the proper models to generate final submission. All proposed framework is implemented in PyTorch using an NVIDIA Tesla V100 GPU.

    \begin{table*}[ht]
		\centering
		\caption{Quantitative experiment of the proposed segmentation framework.}\label{table:t1}
		\renewcommand{\arraystretch}{1.0}
    	\scalebox{0.59}{
    	\Large
		\begin{tabular}{p{7.5cm}<{\centering}|p{3cm}<{\centering}|p{3cm}<{\centering}|p{3cm}<{\centering}|p{3cm}<{\centering}}
			\toprule[2pt]
			%\hline
			\multicolumn{1}{c|}{\bf{Method}}&\multicolumn{1}{c|}{\bf{Task1  DSC}}&\multicolumn{1}{c|}{\bf{Task1 SDSC}}&\multicolumn{1}{c|}{\bf{Task2 DSC}}&\multicolumn{1}{c}{\bf{Task2 SDSC}}\\
			\toprule[1pt]
			\toprule[1pt]
			nnU-Net~\cite{isensee2018nnu} & 0.872 & 0.974 &0.777 &0.926  \\
			ResUnet & 0.873 & 0.973 & 0.780 & 0.929  \\
			ResUnet+Tversky & 0.875 & 0.973 & 0.779 & 0.928  \\
			Ensembled pseudo training  & \textbf{0.876} & \textbf{0.974} & \textbf{0.782} & \textbf{0.929}  \\
			\bottomrule[2pt]
		\end{tabular}}
	\end{table*}

	\subsection{Quantitative and Qualitative Analysis} 
	We evaluate the performance of the proposed frameworks on the online test dataset, with the evaluation metrics as Dice score (DSC) and surface Dice score (SDSC) for Task1 and Task2. The experiment result as shown in Table ~\ref{table:t1}, we choose nnU-Net as the state-of-the-art for comparison, which achieves 87.2\% in DSC and 97.4\% in SDSC of Task1 and 77.7\% in DSC and 92.6\% in SDSC of Task2, respectively. In contrast to nn-UNet, our ResUnet achieves the improvement with 0.3\% in DSC and 0.3\% in SDSC of Task2, and ResUnet with Tversky loss achieves improvement with 0.1\% of DSC and 0.1\% of SDSC in Task2. To further boost the framework performance, we ensemble the prediction of above methods to get better results.
	
\begin{figure}[ht]
	\centering
	\includegraphics[width=0.9\columnwidth]{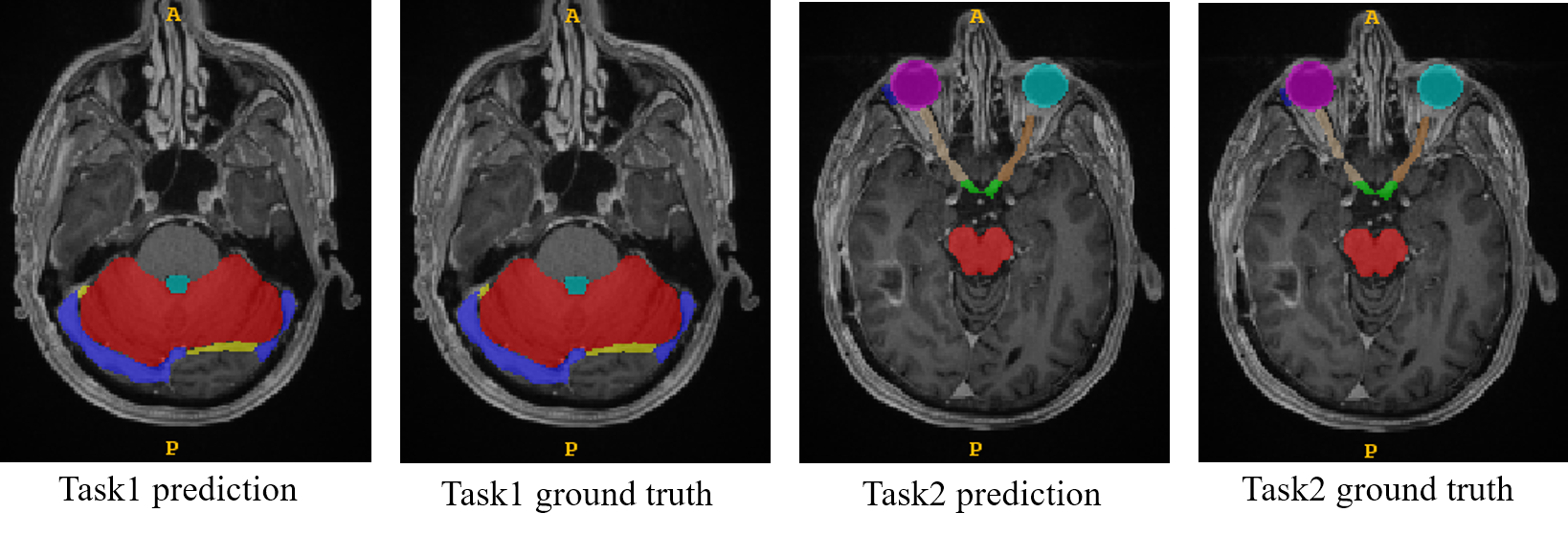}
	\caption{Examples of framework's prediction and corresponding ground truth.}
	\label{fig:results}
    \end{figure}

\begin{figure}[ht]
	\centering
	\includegraphics[width=0.9\columnwidth]{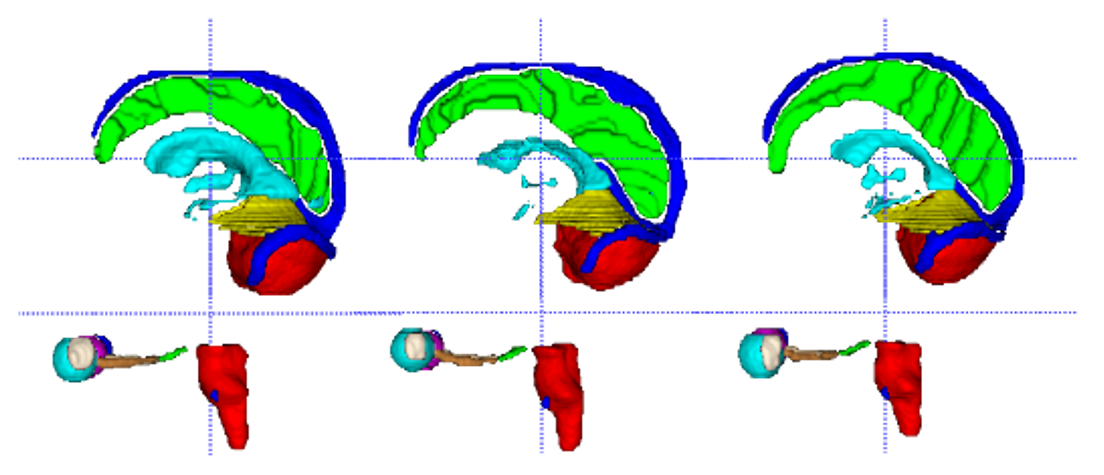}
	\caption{3D reconstructions of framework's prediction.}
	\label{fig:3d}
    \end{figure}

	We visualize the predictions for a qualitative analysis. As shown in  Fig.~\ref{fig:results} the predictions of our proposed framework are very close to corresponding labels. We also reconstructed the prediction in 3D in Fig.~\ref{fig:3d}, where all target issues are clearly identified.

	\section{Conclusions}
    In this study, we proposed a effective framework for Anatomical Brain Barriers to Cancer Spread challenge. Specifically, we used residual block based U-shape network as the proposed architecture and the Tversky loss as the criterion, to enforce the the feature extraction ability. The ensemble strategy was adopted to refine the prediction and get the better result. Experiments on the online test set varied the efficacy of proposed framework.
	\vspace{6mm}

	% ---- Bibliography ----
	%
	\bibliographystyle{splncs}
	\bibliography{ABCs}	
	
\end{document}